# The History of the Grid

Ian Foster[*][+], Carl Kesselman[§][=]

[*]*Computation Institute, Argonne National Laboratory & University of Chicago*
[+]*Department of Computer Science, University of Chicago*
[§]*Department of Industrial and Systems Engineering, University of Southern California*
[=]*Information Sciences Institute, University of Southern California*

**Abstract.** With the widespread availability of high-speed networks, it becomes feasible to outsource computing to remote providers and to federate resources from many locations. Such observations motivated the development, from the mid-1990s onwards, of a range of innovative Grid technologies, applications, and infrastructures. We review the history, current status, and future prospects for Grid computing.

## Introduction

In the 1990s, inspired by the availability of high-speed wide area networks and challenged by the computational requirements of new applications, researchers began to imagine a computing infrastructure that would "provide access to computing on demand" [78] and permit "flexible, secure, coordinated resource sharing among dynamic collections of individuals, institutions, and resources" [81].

This vision was referred to as the Grid [151], by analogy to the electric power grid, which provides access to power on demand, achieves economies of scale by aggregation of supply, and depends on large-scale federation of many suppliers and consumers for its effective operation. The analogy is imperfect, but many people found it inspiring.

Some 15 years later, the Grid more or less exists. We have large-scale commercial providers of computing and storage services, such as

Amazon Web Services and Microsoft Azure. Federated identity services operate, after a fashion at least. International networks spanning hundreds of institutions are used to analyze high energy physics data [82] and to distribute climate simulation data [34]. Not all these developments have occurred in ways anticipated by the Grid pioneers, and certainly much remains to be done; but it is appropriate to document and celebrate this success while also reviewing lessons learned and suggesting directions for future work. We undertake this task in this article, seeking to take stock of what has been achieved as a result of the Grid research agenda and what aspects of that agenda remain important going forward.

**1. A little prehistory**

With the emergence of the Internet, computing can, in principle, be performed anywhere on the planet, and we can access and make use of any information resource anywhere and at any time.

This is by no means a new idea. In 1961, before any effective network existed, McCarthy's experience with the Multics timesharing system led him to hypothesize that "[t]he computing utility could become the basis for a new and important industry" [119]. In 1966, Parkhill produced a prescient book-length analysis [133] of the challenges and opportunities; and in 1969, when UCLA turned on the first node of the ARPANET, Kleinrock claimed that "as [computer networks] grow up and become more sophisticated, we will probably see the spread of 'computer utilities' which, like present electric and telephone utilities, will service individual homes and offices across the country" [106].

Subsequently, we saw the emergence of computer service bureaus and other remote computing approaches, as well as increasingly powerful systems such as FTP and Gopher for accessing remote information. There were also early attempts at leveraging networked computers for computations, such as Condor [112] and Utopia [176]—both still heavily used today, the latter in the form of Platform Computing's Load Sharing Facility [175]. However, it was the emergence of the Web in the 1990s (arguably spurred by the wide availability of PCs with decent graphics and storage) that opened people's eyes to the potential for remote computing. A variety of projects sought to leverage the Web for computing: Charlotte [26],

ParaWeb [38], Popcorn [43], and SuperWeb [9], to name a few. However, none were adopted widely.

The next major impetus for progress was the establishment of high-speed networks such as the US gigabit testbeds. These networks made it feasible to integrate resources at multiple sites, an approach termed "metacomputing" by Catlett and Smarr [45]. Application experiments [122] demonstrated that by assembling unique resources such as vector and parallel supercomputers, new classes of computing resources could be created that were unique in their abilities and customized to the unique requirements of the application at hand [114]. For example, the use of different resource types to execute coupled climate and ocean modeling was demonstrated [120].

Support for developing these types of coupled applications was limited, consisting of network-enabled versions of message-passing tools used for parallel programming [154]. Because these networks were operated in isolation for research purposes only, issues of security and policy enforcement, while considered, were not of primary concern. The promise of these early application experiments led to interest in creating a more structured development and execution platform for distributed applications that could benefit from the dynamic aggregations of diverse resource types. The I-WAY experiment in 1994 [57], which engaged some 50 application groups in demonstrating innovative applications over national research networks, spurred the development of the I-Soft [74] infrastructure, a precursor to both the Globus Toolkit and the National Technology Grid [151]. The book *The Grid: Blueprint for a New Computing Infrastructure* [77] also had a catalyzing effect.

Meanwhile, scientific communities were starting to look seriously at Grid computing as a solution to resource federation problems. For example, high energy physicists designing the Large Hadron Collider (LHC) realized that they needed to federate computing systems at hundreds of sites if they were to analyze the many petabytes of data to be produced by LHC experiments. Thus they launched the EU DataGrid project in Europe [42] and the Particle Physics Data Grid (ppdg.net) and Grid Physics Network [24] projects in the US, two efforts that ultimately led to the creation of the Open Science Grid in the US, EGEE and then EGI in Europe, and the international LHC Computing Grid (LCG) [109]. Figure 1 shows a representative sample of these significant events in Grid development.

Much early work in Grid focused on the potential for a new class of infrastructure that the Grid represented. However, the computing world today looks significantly different now from what it did at the start of the

"Grid era" in ways that transcend simply bigger, faster, and better. Grid computing started at a time when application portability remained a major challenge: many processor architectures competed for dominance, the Unix wars were still raging, and virtualization had not yet emerged as a commodity technology. CORBA was in its ascendency, and Web technology was restricted to basic HTML with `blink` tags, HTTP, and CGI scripts. Today, we have fewer operating systems to support and, with the triumph of x86, fewer hardware platforms. High-quality virtualization support is widely available. The number of implementation languages and hosting environments has grown, but powerful client-side application platforms exist, and there is increasing consolidation around RESTful architectural principles [66] at the expense of more complex Web Services interfaces. Such advances have considerable implications for how today's Grid will evolve.

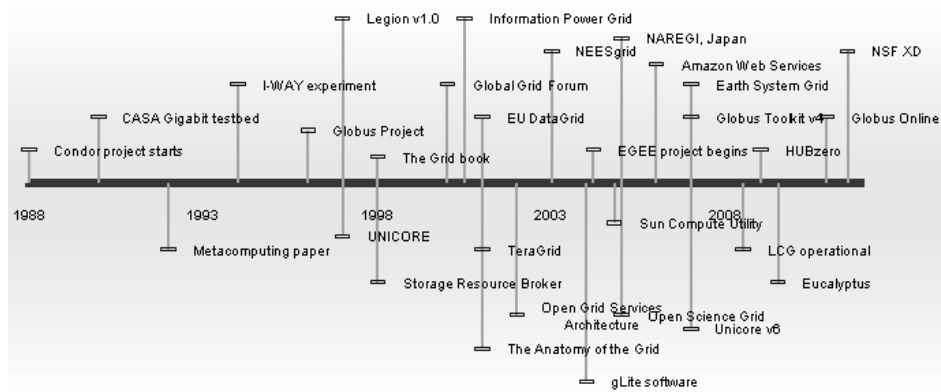

Figure 1: Abbreviated Grid timeline, showing 30 representative events during the period 1988–2011

## 2. Terminology

Any discussion of the Grid is complicated by the great diversity of problems and systems to which the term "Grid" has been applied. We find references to computational Grid, data Grid, knowledge Grid, discovery Grid, desktop Grid, cluster Grid, enterprise Grid, global Grid, and many others. All such systems seek to integrate multiple resources into more powerful aggregated services, but they differ greatly in many dimensions.

One of us defined a three-point checklist for identifying a Grid, which we characterized as a system that does the following [71]:

1. *"[C]oordinates resources that are not subject to centralized control ...* (A Grid integrates and coordinates resources and users that live within different control domains—for example, the user's desktop vs. central computing; different administrative units of the same company; or different companies; and addresses the issues of security, policy, payment, membership, and so forth that arise in these settings. Otherwise, we are dealing with a local management system.)
2. *... using standard, open, general-purpose protocols and interfaces ...* (A Grid is built from multi-purpose protocols and interfaces that address such fundamental issues as authentication, authorization, resource discovery, and resource access. ... [I]t is important that these protocols and interfaces be *standard* and *open*. Otherwise, we are dealing with an application-specific system.)
3. *... to deliver nontrivial qualities of service.* (A Grid allows its constituent resources to be used in a coordinated fashion to deliver various qualities of service, relating for example to response time, throughput, availability, and security, and/or co-allocation of multiple resource types to meet complex user demands, so that the utility of the combined system is significantly greater than that of the sum of its parts.)"

We still think that this checklist is useful, but we admit that no current system fulfills all three criteria. The most ambitious Grid deployments, such as the LHC Computing Grid, Open Science Grid, and TeraGrid, certainly integrate many resources without any single central point of control and make heavy use of open protocols, but they provide only limited assurances with respect to quality of service. The most impressive Grid-like systems in terms of qualities of service—systems like Amazon Web Services—coordinate many resources but do not span administrative domains. So perhaps our definition is too stringent.

In the rest of this section, we discuss briefly some of the infrastructures to which the term Grid has been applied.

The term **computational Grid** is often used to indicate a distributed resource management infrastructure that focuses on coordinated access to remote computing resources [76]. The resources that are integrated by such infrastructures are typically dedicated computational platforms, either high-end supercomputers or general-purpose clusters. Examples include the US TeraGrid and Open Science Grid and, in Europe, the UK National Grid Service, German D-Grid, INFN Grid, and NorduGrid.

Grid functions, which are primarily about resource aggregation and coordinated computation management, often have been confused with local resource managers [86], such as the Portable Batch System (PBS), Load Sharing Facility (LSF) [175], and Grid Engine [87], whose function is limited to scheduling jobs to local computational nodes in a manner that is consistent with local policy. Complicating the picture is the issue that many local resource managers also incorporate mechanisms for distributed resource management, although these functions tend to be limited to scheduling across resources within an enterprise [86].

The emergence of infrastructure-as-a-service (IaaS) providers [121] such as Amazon EC2 and Microsoft Azure are sometimes assumed to solve the basic needs of computational Grid infrastructure. But these solutions are really alternatives to local resource management systems; the issues of cross-domain resource coordination that are at the core of the Grid agenda remain. Indeed, the cloud community is starting to discuss the need for "intercloud protocols" and other concepts familiar within Grids, and cloud vendors are starting to explore the hierarchical scheduling approaches ("glide-ins") that have long been used effectively in Grid platforms.

**Desktop Grids** are concerned with mapping collections of loosely coupled computational tasks to nondedicated resources, typically an individual's desktop machine. The motivation behind these infrastructures is that unused desktop cycles represented potentially enormous quantities (ultimately, petaflops) of computing. Two distinct usage models have emerged for such systems, which David Anderson, a pioneer in this space, terms (somewhat confusingly, given our desktop Grid heading) *volunteer* and *grid* systems, respectively. (Desktop grids have also been referred to as *distributed* [108] and *peer-to-peer* [124] computing.) In the former case, volunteers contribute resources (often home computers) to advance research on problems that often have broad societal importance [17], such as drug discovery, climate modeling, and analyzing radio telescope data for evidence of signals (SETI@home [16]). Volunteer computing systems must be able to deal with computers that are often unreliable and poorly connected. Furthermore, because volunteer computers cannot be trusted, applications must be resilient to incorrect answers. Nevertheless, such systems—many of which build on the BOINC [15] platform—often deliver large quantities of computing. XtremWeb [65] is another infrastructure created for such computing.

The second class of desktop Grids deployments occurs within more controlled environments, such as universities, enterprises, and individual research projects, in which participants form part of a single organization (in which case, we are arguably not dealing with a Grid but, rather, a local resource manager) or virtual organization. In these settings, Condor [112] has long been a dominant technology.

Some authors have characterized federated data management services as forming a **data Grid** [46, 141]. This terminology is somewhat unfortunate in that it can suggest that data management requires a distinct Grid infrastructure, which is not the case. In reality, data often needs to be analyzed as well as managed, in which case data management services must be combined with computing, for example to construct data analysis pipelines [83]. With this caveat, we note that various systems have been developed that are designed primarily to enable the federation and management of (often large) data sets: for example, the LIGO Data Grid [5], used to distribute data from the Laser Interferometer Gravitational Wave Observatory (LIGO) [28] to collaboration sites in Europe and the US; the Earth System Grid [34], used to distribute climate data to researchers worldwide; and the Biomedical Informatics Research Network (BIRN) [95].

Peer-to-peer file sharing systems such as BitTorrent [51] have also created large-scale infrastructures for reliable data sharing. While responsible for significant fractions of Internet traffic, their design points with respect to security and policy enforcement (specifically, the lack of either) are significantly different from those associated with Grid applications and infrastructure.

The term **service Grid** is sometimes used to denote infrastructures that federate collections of application-specific Web Services [37], each of which encapsulates some data source or computational function. Examples include virtual observatories in astronomy [156], the myGrid [152] tools for federating biological data, the caGrid infrastructure in cancer research [131], and the Cardio Vascular Research Grid (CVRG) [1]. These systems combine commodity Web Services and (in some cases) Grid security federation technologies to enable secure sharing across institutional boundaries [70].

## 3. Grid lifecycle

To understand how Grids have been created and operationed, let us consider the power grid analogy introduced in Section 1 and examine the correspondence between the power grid and the computational Grids that we study here. We observe that while the electric infrastructures are public utilities, customer/provider relationships are well defined. We also observe that co-generation issues aside, the infrastructure by which power utilities share resources (power) is governed by carefully crafted business relationships between power companies.

In many respects, the way in which Grid infrastructure has been built, deployed, and operated mirror these structures. Grid infrastructure has not formed spontaneously but rather is the result of a deliberate sequence of coordinated steps and (painfully) negotiated resource-sharing agreements. These steps have tended to be driven by dedicated operational teams. This model has been followed in virtually all major Grid deployments, including Open Science Grid, TeraGrid, the NASA Information Power Grid, various other national Grids, and LCG. More organic formulation of Grid infrastructure has been limited by the complexities of the policy issues, the difficulty in dynamically negotiating service level agreements, and, until recently, the lack of a charging model.

Looking across a number of operational Grid deployments, we identify the following common steps in the lifecycle of creating, deploying, and operating a Grid infrastructure:

1. **Provisioning resources/services to be made available**. Resource owners allocate, or *provision,* existing or newly acquired computing or storage systems for access as part of a federated Grid infrastructure. This work may involve setting up dedicated submission queues to a batch-scheduled resource, creating Grid user accounts, and/or altering resource usage policy. In research settings, the resources accessible for the Grid are often not purchased explicitly for that purpose, and Grid usage must be balanced against local community needs. The emergence of for-profit IaaS providers offers the potential for more hands-off provisioning of resources and has greatly streamlined this process.
2. **Publishing those resources** by making them accessible via standardized, interoperable network interfaces (protocols). In many production Grids, the Globus Toolkit components such as GRAM and GridFTP provided these publication mechanisms by supplying standardized network interfaces by which provisioned resources can be used in wide area, multisite settings. Other widely used publication interfaces include Unicore [143, 153] and the Basic Execution Services (BES) [75] defined within the Open Grid Forum.

3. **Assembling the resources into an operational Grid.** The initial vision for the Grids was dynamic assembly of interoperable resources. The most successful production Grids, however, have involved the careful integration of resources into a common framework, not only of software, but also of configuration, operational procedures, and policies. As part of this collection, operational teams define and operate Grid-wide services for functions such as operation, service discovery, and scheduling. Furthermore, production Grids have typically required substantial software stacks, which necessitated complex software packaging, integration, and distribution mechanisms. An unfortunate consequence of this part of the Grid lifecycle was that while these Grids achieved operability between independently owned and operated resources, interoperability between production Grid deployments was limited. Viewed from this perspective, production Grids have many characteristics in common with IaaS providers.
4. **Consuming those resources through a variety of applications.** User applications typically invoke services provided by Grid resource providers to launch application programs to run on computers within the Grid; to carry out other activities such as resource discovery and data access; or to invoke software for which a service interface is provided. User interactions with the Grid may involve the use of thick or thin clients and are often facilitated by client libraries that encapsulate Grid service operations (e.g., COG Kit [165]).

## 4. Applications

Work on applications has been motivated by the availability of infrastructure and software and has, in turn, driven the development of that infrastructure and software. We review here some important classes of Grid applications (see also [52]).

Interest in Grid computing has often been motivated by applications that invoke many independent or loosely coupled computations. Such applications arise, for example, when searching for a suitable design, characterizing uncertainty, understanding a parameter space [7], analyzing large quantities of data, or engaging in numerical optimization [20, 162]. Scheduling such loosely coupled compute jobs onto Grid resources has proven highly successful in many settings. Such applications are malleable to the changing shape of the underlying resources and can often be structured to have limited data movement requirements. They are the mainstay of Grid environments operated by the high energy and nuclear physics community, including the Open Science Grid and the LCG. High-throughput [113] or many-task [139] computations require large amounts of computing, which Grid infrastructures can often provide at modest cost. Such applications have in turn motivated the development of specialized schedulers and job managers (e.g., Condor [112], Condor-G [84]) and new programming models and tools variously referred to as parallel scripting [172] and workflow [56, 157].

Tightly coupled applications are less commonly executed across multiple Grid-connected systems; more commonly, Grid systems are used to dispatch such applications to a single remote computer for execution. However, several projects have sought to harness multiple high-end computer systems for such applications. Adaptations such as clever problem decompositions or approximation methods at various points in a simulation may be used to reduce communication requirements. An early experiment in this area was SF-Express, a "synthetic forces" discrete event simulation application that coupled large compute clusters at multiple sites to simulate collections of more than 100,000 entities [39]. A number of other such applications have been developed [13, 116, 123], including impressive large-scale fluid dynamics and other computational physics simulations [33, 58, 116]. However, the fundamental conflict between resource providers and consumers for anything but best effort service means that such experiments have involved mostly one-off demonstrations. While resource reservation methods [55, 73] and associated co-allocation algorithms [54, 115] have been explored, these coordination models have not seen wide adoption because of the cost and complexity of reserving expensive and generally oversubscribed resources.

Other important Grid applications have involved the remote operation of, and/or analysis of data from, scientific instrumentation [99, 100, 135, 167] or other devices [111]. A related set of applications has focused on the distribution and sharing of large amounts of digital content—for example, digital media [94], gravitational wave astronomy data [47], and medical images [14, 62]. Biomedical applications have emerged as a major driver of Grid computing, because of their need to federate data from many sources and to perform large-scale computing on that data [61, 117, 149]. Opportunities appear particularly large in so-called translational research [145].

A different class of Grid applications focused on the incorporation of multimedia data such as sound and video to create rich, distributed collaboration environments. For example, Access Grid [50] uses a variety of Grid protocols to create virtual collaboration spaces including immersive audio and video. The social informatics data Grid (SIDgrid) [35] built on Access Grid to create distributed data repositories that include not only numerical, text, and image data but also video and audio data, in order to support social and behavioral research that relies on rich, multimodal behavioral information.

**5. Grid architecture, protocols, and software**

The complexities inherent in integrating distributed resources of different types and located within distinct administrative domains led to a great deal of attention to issues of architecture and remote access protocols and to the development of software designed variously to mask and/or enable management of various heterogeneities.

Figure 2 shows a commonly used depiction of Grid architecture, from 1999 [81]. In brief, the Fabric comprises the resources to which remote access is desired, while Connectivity protocols (invariably Internet Protocol based) permit secure remote access. Resource layer protocols enable remote access to, and management of, specific classes of resources; here we see modeling of, for example, computing and storage resources. Collective services and associated protocols provide integrated (often virtual organization-specific: see Section 7) views of many resources.

*5.1. Grid middleware*

With the transition from small-scale, experimental gigabit wide-area networks to more persistent national and international high-speed backbones, the need for less ad hoc methods for coordinating and managing multiresource applications became pressing. Issues that needed to be addressed included allocating and initiating cross-site computations on a range of different computing platforms, managing executables, providing access to program outputs, communicating between program components, monitoring and controlling ongoing computations, and providing cross-site authentication and authorization. These requirements resulted in the development and evaluation of a range of different infrastructure solutions. Strategies investigated included distributed-memory approaches, leveraging of ancient Web server technology, distributed object systems, remote procedure call systems, and network services architectures. We highlight a few of the more prominent solutions below.

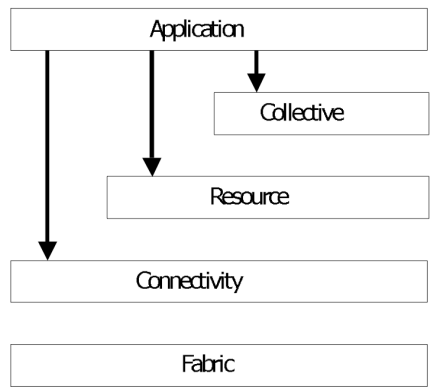

Figure 2: Simple view of Grid architecture: see text for details

When Grid work started, no good methods existed for publishing and accessing services. Distributed Computing Environment (DCE) [6] and Common Object Request Broker Architecture (CORBA) [130] were available but were oriented toward more tightly coupled and controlled enterprise environments. While some attempts were made to adapt these technologies to Grid environments, such as the Common Component Architecture [21], the high level of coordination generally required to deploy and operate these infrastructures limited their use.

The emergence of loosely coupled service-oriented architectures was of great interest to the Grid community. Initial focus was on SOAP-based Web Services. This effort comprised two aspects. One was the use of the tooling and encodings that SOAP provided. There were also rich, layered sets of additional standards and components that layered on top of these basic interfaces for security, various transports, and so forth. The second aspect of this effort was the more explicit adoption of so-called service-oriented architectures as an underlying architectural foundation. Grid projects such as Globus were early adopters of these approaches. Tools were immature. Additional layering caused performance issues. Some infrastructures such as Condor never adopted these technologies. The Globus Toolkit followed a hybrid approach with "legacy" interfaces (e.g., GRAM) supported alongside newer SOAP interfaces.

Legion [92] was an early Grid infrastructure system. Its system model was a consistent object-oriented framework. It used public keys for authentication and provided a distributed file system abstraction and an object-oriented process invocation framework. The Legion system is no longer in use.

An alternative approach was taken by Unicore [143, 153], which was developed by a consortium of university and industry partners. The central idea behind Unicore is to provide a uniform job submission interface across a wide range of different underlying job submission and batch management systems. Unicore is architected around a modular service-oriented architecture and is still in active development, being used, for example, in the large-scale European Grid infrastructure projects DEISA [88] and PRACE [23].

Perhaps the best-known and most widely deployed Grid middleware infrastructure is the Globus Toolkit. Globus is architected around an Internet-style hourglass architecture and consists of an orthogonal collection of critical services and associated interfaces. Key components include the use of X.509 proxy certificates for authentication and access control, a layered monitoring architecture (MDS), a HTTP-based job submission protocol (GRAM), and a high-performance data management service based on FTP (GridFTP). Globus has served as the foundation of most Grid infrastructures deployed outside Europe and also plays a significant role in European infrastructure deployments, including ARC [59], gLite [110], and DEISA [74], although those systems certainly also include substantial other components. In addition, Globus serves as the foundation of other Grid infrastructure toolkits, such as the National Institutes of Health caGrid infrastructure [131] that underpins the cancer Biomedical Informatics Grid (caBIG).

Many task computations frequently use a two-level scheduling approach, in which a Grid-based resource management protocol such as GRAM is used to deploy, or *glide in* [147], higher-level application environments, such as Condor scheduling services [32, 158]. This approach allows Grid infrastructure to act in much the same way as current cloud-based IaaS providers.

*5.2. Data management middleware*

Management of computing resources has tended to be a core component of all Grid middleware. However, the inevitable increase in the amount of data generated driven by ever more detailed and powerful simulation models and scientific instruments led to the creation of Grid services for managing multiterabyte datasets consisting of hundreds of thousands or millions of files. At one extreme, we saw large-scale physical simulations that could generate multigigabyte data files that captured the simulation state at a given point in time, while at the other extreme we saw applications such as those in high energy physics that would generate millions of smaller files. (With a few notable exceptions, such as the SkyServer work done by Szalay and Gray as part of the National Virtual Observatory [13], most Grid data management systems dealt with data almost exclusively at the level of files, a tendency critiqued by Nieto-Santisteban et al. [127]. )

Many different data management solutions have been developed over the years for Grid infrastructure. We consider three representative points in the solution space. At the most granular end of the spectrum is GridFTP, a standardized extension of the FTP protocol [11], that provides a robust, secure, high-performance file transfer solution that performs extremely well with large files over high-performance networks. One important feature is its support for third-party transfer, enabling a hosted application to orchestrate data movement between two storage endpoints. GridFTP has seen extensive use as a core data mover in many Grid deployments, with multiple implementations and many servers in operation. Globus GridFTP [10] and other data management services, such as its Replica Location Service [48], have been integrated to produce a range of application-specific data management solutions, such as those used by the LIGO Data Grid [5], Earth System Grid [34], and QCDgrid [136]. The more recent Globus Online system builds on Globus components to provide higher-level, user-facing, hosted research data management functions [12, 68].

Higher levels of data abstraction were provided by more generic data access services such as the OGSA Data Access and Integration Service developed at EPCC at the University of Edinburgh [22]. Rather than limiting data operations to opaque file containers, OGSA-DAI enables access to structured data, including structured files, XML representations, and databases. DAI achieves this by providing standard Grid-based read and write interfaces coupled with highly extensible data transformation workflows called *activities* that enable federation of diverse data sources. A distributed query processor enables distributed, Grid-based data sources to be queried as a single virtual data repository.

At the highest level of abstraction are complete data management solutions that tend to focus on data federation and discovery. For example, the Storage Resource Broker [13][30] and the follow-on Integrated Rule-Oriented Data System [140] facilitate the complete data management lifecycle: data discovery via consolidated metadata catalogs, policy enforcement, and movement and management, including replication for performance and reliability as well as data retrieval.

*5.3. Grid application software*

One common approach to supporting the creation of Grid applications was the creation of versions of common parallel programming tools, such as MPI, that operated seamlessly in a distributed, multiresource Grid execution environment [60]. An example of such a tool is MPICH-G [103] (now MPIg), a Globus-enabled version of the popular MPICH programming library. MPICH-G uses job coordination features of GRAM submissions to create and configure MPI communicators over multiple co-allocated resources and configures underlying communication methods for efficient point-to-point and collective communications. MPICH-G has been used to run a number of large-scale distributed computations.

Another common approach to providing Grid-based programming environments is to embed Grid operations for resource management, communication, and data access into popular programming environments. Examples include pyGlobus [96] and the Java COG Kit [165], both of which provide object-based abstractions of underlying Grid abstractions provided by the Globus toolkit. A slightly different approach was taken in the Grid Application Toolkit (GAT) [146] and its successor, the Simple API for Grid Applications (SAGA) [97], both of which seek to simplify Grid programming in a variety of programming languages by providing a higher-level interface to basic Grid operations.

What have been variously termed portals [159], gateways [173], and HUBs emerged as another important class of Grid application enablers. Examples include the UCLA Grid portal, GridPort [159], Hotpage [160], the Open Grid Computing Environment [8], myGrid [91], and nanoHUB [107]. Focusing on enabling broad community access to advanced computational capabilities, these systems have variously provided access to computers, applications, data, scientific instruments, and other capabilities. Remote job submission and management are a central function of these systems. Many special-purpose portals have been created for this use and have seen (and continue to see) widespread use in centers that operate capability resources.

*5.4. Security technologies*

In the early days of Grid computing, security was viewed as a major roadblock to the deployment and operation of Grid infrastructure. (Recall that in the early 1990s, plaintext passwords were still widely used for authentication to remote sites.) Such concerns spurred a vigorous and productive R&D program that has produced a robust security infrastructure for Grid systems. This R&D program has both borrowed from and contributed to the security technologies that underpin today's Internet. One measure of its success is that in practice, most major Grid deployments have used open Internet connections rather than private networks or virtual private networks (VPNs), as many feared would be required in the early days of the Grid.

One early area of R&D focus concerned the methods to be used for mutual authentication of users and resources and for subsequent authorization of resources access. In the early 1990s, Kerberos [126] was advocated (and used) by some as a basis for Grid infrastructures [31]. However, concerns about its need for interinstitutional agreements led to adoption of public key technology instead [40]. The need for Grid computations to delegate authority [85] to third parties, as when a user launches a computation that then accesses resources on the user's behalf, led to the design of the widely adopted Grid Security Infrastructure [80] and its extended X.509 proxy certificates [163, 169]. These concepts and technologies still underpin today's Grid, but they have been refined greatly over time.

In the first Grid systems, authorization was handled by GridMap files (a simple form of access control list) associated with resources. While simple, this approach made basic tasks such as adding a new user to a collaboration a challenge, requiring updates to GridMap files at many locations. The Virtual Organization Management Service (VOMS) [64] has been widely adopted as a partial solution to this problem. (The Community Authorization Service [134] was another early system.) The Akenti system [161] pioneered attribute-based authorization methods that, in more modern forms, have been widely adopted [170]. Meanwhile, security technologies were integrated into commonly used libraries for use in client applications Welch [171]

The need for users to manage their own X.509 credentials proved to be a major obstacle to adoption and also a potential vulnerability. One partial solution was the development of the MyProxy online credential repository [128]. The use of online Certification Authorities integrated with campus authorization infrastructures (e.g., via Shib [63]) means that few Grid users manage their own credentials today [168]. Integration with OpenID has also been undertaken [148].

*5.5. Portability concerns*

Application portability is perhaps the significant obstacle to effective sharing of distributed computational resources. The increased adoption of Linux as an operating system for scaleout computing platforms resolved a number of the more significant portability issues. Careful use of C and Fortran programming libraries along with the advent of Java further addressed portability issues. However, variations in the configuration of local system environments such as file systems and local job management system continued (and, indeed, continue today) to complicate the portability of jobs between Grid nodes.

The standardized job submission and management interfaces provided by Grid infrastructures such as GRAM and DRMAA [142] simplified the task of providing site independence and interoperability. However, local configuration details, such as file system locations, different versions of dynamically linked libraries, scheduler idiosyncrasies, and storage system topologies, tended to restrict scheduling flexibility. Within Grid deployments, several simple mechanisms have proven useful, such as requiring participating resource providers to set a minimal set of environment variables [44], standardizing configurations of compute and storage nodes, and the use of federated namespaces, such as global file systems.

At the application level, systems such as Condor helped ameliorate these portability issues by trapping and redirecting environment-specific operations, such as file creation, to a centralized server. Nevertheless, true independence of computational tasks remains a difficult process, and we see limited portability of programs between Grid platforms.

Recent advances in both the performance and the ubiquity of virtual machine technology have significantly improved application portability, while also providing security benefits [67]. However, differences in hypervisor environments and Linux distributions mean that truly portable scheduling of virtual machines across a Grid of cloud platforms is still not a solved problem.

## 6. Infrastructures

The past decade has seen the creation of many Grid infrastructure deployments. Some of the earliest large-scale deployments were organized programmatically to support targeted user communities. Perhaps the first was NASA's Information Power Grid (IPG) [101], designed to integrate the various supercomputer centers at NASA laboratories into an integrated computing framework. Based primarily on the Globus Toolkit, the IPG program was responsible for identifying many of the critical operational issues of Grid infrastructure around monitoring, user support, application development, and global research management. Other examples include Grids to support high energy and nuclear physics (e.g., LCG – see Figure 3, Open Science Grid), climate research (e.g., Earth System Grid [34]), earthquake engineering research [105], and gravitational wave astronomy [28]. The Dutch-distributed ASCI supercomputer [25] and the French Grid5000 system [36] have both enabled a broad range of innovative computer science research. (In the US, FutureGrid [166] seeks to fill a similar role.)

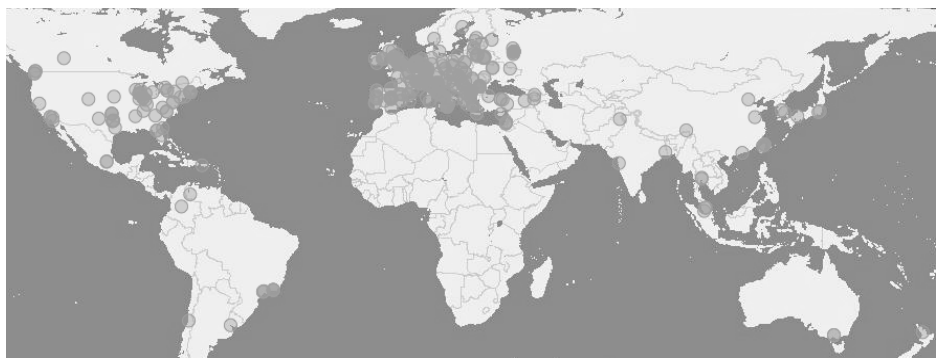

Figure 3: LHC Computing Grid sites as of June 2011 (from http://gstat-prod.cern.ch)

Many of these efforts have been coordinated and financed as national-scale efforts to support the scientific research community within a country. Examples of such deployments include ChinaGrid [98], the UK's National Grid Service (ngs.ac.uk), the Broadband-enabled Science and Technology Grid (BeSTgrid) in New Zealand (bestgrid.org) [102], Australia, ThaiGrid in Thailand (thaigrid.or.th) [164], German D-Grid (dgrid.de) [89], INFNgrid (italiangrid.org), DutchGrid (dutchgrid.nl) and Distributed ASCI Supercomputer (DAS) in the Netherlands, NorduGrid (nordugrid.org) in the Nordic countries [59], Garuda Grid in India [138], NAREGI in Japan [118], and the Open Science Grid in the US [137].

Building on national Grid infrastructures, a number of international Grid deployments were developed, such as the European Union DataGrid [42] and its follow-ons, EGEE and the European Grid Infrastructure (EGI: egi.eu). Several of these Grids, such as the Open Science Grid and NorduGrid, use the Virtual Organization concept [81] (see next section) as a central organizing principle.

## 7. Services for Virtual Organizations

One of the most important features of Grid infrastructures, applications, and technologies has been an emphasis on resource sharing within a *virtual organization*: a set of individuals and/or institutions united by some common interest, and working within a virtual infrastructure characterized by rules that define "clearly and carefully just what is shared, who is allowed to share, and the conditions under which sharing occurs" [81]. This term was introduced to Grid computing in a 1999 article [81], although it previously had been used in the organizational theory literature to indicate purely human organizations, such as inter-company distributed teams [125, 132].

The virtual organization as an organizing principle emphasizes the use of Grid technologies to enable resource federation rather than just on-demand supply. Particularly within the world of science, resource-sharing relationships are fundamental to progress, whether concerned with data (e.g., observational and simulation data within the climate community [34], genome and clinical data within biomedicine), computers (e.g., the international LCG used to analyze data from the Large Hadron Collider), or scientific instrumentation. Such sharing relationships may be long-lived (e.g., the LHC is a multidecade experiment) or short-lived (e.g., a handful of researchers collaborate on a paper, or on a multisite clinical trial); see Figure 4.

The virtual organization (VO) places challenging demands on computing technologies. A set of individuals, who perhaps have no prior trust relationships, need to be able to establish trust relationships, describe and access shared resources, and define and enforce policies concerning who can access what resources and under what conditions. They may also want to establish VO-specific collective services (see Section 5) for use by VO participants, such as group management services; directory services for discovering and determining the status of VO resources and services [53]; metascheduling services for mapping computational tasks to computers; data replication services to keep data synchronized across different collaborating sites; and federated query services. In effect, they need to instantiate at least some fraction of the services that define a physical organization, and to manage and control access to those services much as a physical organization would do.

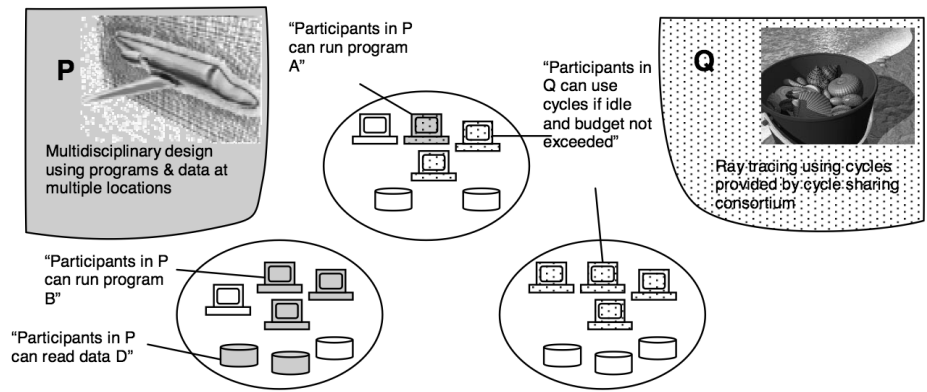

Figure 4: An actual organization can participate in one or more VOs by sharing some or all of its resources. We show three actual organizations (the ovals) and two VOs: P, which links participants in an aerospace design consortium, and Q, which links colleagues who have agreed to share spare computing cycles, for example to run ray tracing computations. The organization on the left participates in P, the one to the right participates in Q, and the third is a member of both P and Q. The policies governing access to resources (summarized in "quotes") vary according to the organizations, resources, and VOs involved. (From [81].)

In principle, the instantiation of a VO could and should be a lightweight operation, and VOs would be created, modified, and destroyed frequently. In practice, VO management tasks remain fairly heavyweight, because many relevant activities are performed manually rather than automatically. Nevertheless, technologies such as the Virtual Organization Management Service (VOMS) [64] and Grouper [3], as well as authorization callouts incorporated into Grid infrastructure services such as GRAM and GridFTP, are gradually reducing the cost of managing distributed VO infrastructures.

## 8. Adventures with standards

Recognizing the success of the Internet standards in federating networks and the fact that Grids were about resource sharing and federation, the Grid community realized the need for standardization early on. Thus in 1999, Ian Foster and Bill Johnston convened the first meeting, at NASA Ames Research Center, of what eventually became the Grid Forum (and later the Global Grid Forum and then the Open Grid Forum (OGF), as a result of mergers with other organizations). Charlie Catlett served as the first chair of these organizations.

Success in the standards space can be measured by two independent metrics: the extent to which an appropriate, representative, and significant subset of the community agree on the technical content; and, given technical content, the extent to which there is appreciable (and interoperable) implementation and deployment of those standards.

Work in OGF and elsewhere (IETF, OASIS) led to successful standards along both these dimensions, notably the proxy certificate profile [163] that underpins the Grid Security Infrastructure [80] and the GridFTP extensions [11] to the File Transfer Protocol—both of which are widely used, primarily in Grid infrastructures targeted to science and research. Other efforts that have enabled substantial interoperation include the Storage Resource Manager specification [93] and the policy specifications that underpin the International Grid Trust Federation (www.igtf.net). The Grid Laboratory for a Uniform Environment (GLUE) specification [18] has facilitated the federation of task execution systems, for example within the high energy physics community.

Other standardization efforts were less successful in terms of wide-scale adoption and use. A substantial effort involving multiple industry and academic participants produced first the Open Grid Services Infrastructure (OGSI) [79] and then the Web Services Resource Framework (WSRF) [72]. The intention was to capture, in terms of a small set of Web Services operations, interaction patterns encountered frequently in Grid deployments, such as publication and discovery of resource properties, and management of resource lifetimes. These specifications were implemented by several groups, including Globus 4 [69], Unicore [116], and software venders, and used in substantial applications. However, inadequate support within Web Services integrated development environments, industry politics, and some level of disappointment with Web Services have hindered widespread adoption. In retrospect, these specifications were too ambitious, requiring buy-in from too many people for success. We expect these specifications to disappear within the next few years.

Many other Grid standardization efforts have been undertaken at higher levels in the software stack. A major impetus for such efforts appears often to have been encouragement from European funding agencies, perhaps on the grounds that this is a good way to influence the computing industry in a way that meets European interests. However, these efforts have not necessarily had the desired effect. Researchers have spent much time developing specifications not because of a need to interoperate (surely the only compelling reason for standardization) but because their research contract required them to do so. As a result, many recent OGF specifications address esoteric issues and have not seen significant adoption.

One area in which the jury remains out is execution service specifications. The importance of resource management systems in many areas of computing has led to frequent calls for standardization. For example, Platform Computing launched in 2000 its New Productivity Initiative to develop a standard API for distributed resource management. In 2002, this effort merged with similar efforts in OGF. One outcome was the DRMAA specification [115]; more recent efforts have produced the Job Submission Description Language (JSDL) [19] and Basic Execution Service (BES) [75] specifications. These specifications fulfill useful functions; however, while deployed in a number of infrastructures, for example in European Union-funded projects, they have yet to make a significant impact. Meanwhile, attention in industry has shifted to the world of cloud computing, where standards are also needed—but seem unlikely for the moment, given the commercial forces at work. It is unclear where these efforts will lead.

Perhaps a fundamental issue affecting the differential adoption of different Grid standards is that while people often want to share data (the focus of Internet protocols and GridFTP, for example), they less frequently want to share raw computing resources. With the increased uptake of software as a service, the distinction between data and computing is blurring, further diminishing the role of execution interfaces.

## 9. Commercial activities

The late 1990s saw widespread enthusiasm for Grid computing in industry. Many vendors saw a need for a Grid product. Because few had on-demand computing or resource federation capabilities to offer, many cluster computing products became Grid products overnight. (One vendor's humble blade server became a Grid server; 10 years later it was to become a cloud server.) We review a few of these commercial offerings here.

One early focus of commercial interest was the desktop Grid. Entropia [49] and United Devices were two of several companies that launched products in this space. Their goal initially was to monetize access to volunteer computers, but they found few customers because of concerns with security, payback, and limited data bandwidth. Both saw more success in enterprise deployments where there was more control over desktop resources; some industries with particularly large computing needs, such as the financial services industry, became heavy users of desktop Grid products. Today, this market has mostly disappeared, perhaps because the majority of enterprise computing capacity is no longer sitting on employee desktops.

Another set of companies targeted a similar market segment but using what would have been called, prior to the emergence of Grid, resource management systems. Platform with its Load Sharing Facility and Sun with its Grid Engine both sought to deliver what they called "cluster Grid" or "enterprise Grid" solutions. Oracle gave the Grid name to a parallel computing product, labeling their next-generation database system Oracle 10G. The "G" in this case indicated that this system was designed to run on multiprocessors. This restricted view of Grid computing has become widespread in some industries, leading one analyst to suggest that "if you own it, it's a grid; if you don't, it's a cloud." Contrast this view with that presented in Section 2.

Several startup companies, ultimately unsuccessful, sought to establish computational markets to connect people requiring computational capacity with sites with a momentary excess of such capacity. (This same concept has also generated much interest in academia [41, 43, 155, 174].) These people saw, correctly, that without the ability to pay for computational resources, the positive returns to scale required for large-scale adoption of Grid technology could not be achieved. However, they have so far been proven incorrect in their assumption that a monetized Grid would feature many suppliers and thus require market-based mechanisms to determine the value of computing resources. Instead, today's cloud features a small number of suppliers who deliver computing resources at fixed per unit costs. (However, Amazon has recently introduced a "spot market" for unused cycles.)

Few if any companies made a business out of Grid in the traditional sense. Univa Corporation was initially founded to support the Globus Toolkit; its product line has evolved to focus on open source resource management stacks based on Grid Engine. Avaki ultimately failed to create a business around distributed data products (Data Grid). IBM created a product offering around Grid infrastructure, leveraging both the Globus Toolkit and their significant investment in SOAP-based Web Services. They had some success but did not create a huge business.

We attribute the lackluster record of commercial Grid (outside some narrow business segments such as financial services), relative to its widespread adoption in science, to two factors. First, while resource sharing is fundamental to much of science, it is less frequently important in industry. (One exception to this statement is large, especially multinational, companies—and they were indeed often early adopters.) Second, when commercial Grid started, there were no large suppliers of on-demand computational services—no power plants, in effect. This gap was filled by the emergence of commercial infrastructure as a service ("cloud") providers, as we discuss next.

## 10. Cloud computing

The emergence of cloud computing around 2006 is a fascinating story of marketing, business model, and technological innovation. A cynic could observe, with some degree of truth, that many articles from the 1990s and early 2000s on Grid computing could be—and often were—republished by replacing every occurrence of "Grid" with "cloud." But this is more a comment on the fashion- and hype-driven nature of technology journalism (and, we fear, much academic research in computer science) than on cloud itself. In practice, cloud is about the effective realization of the economies of scale to which early Grid work aspired but did not achieve because of inadequate supply and demand. The success of cloud is due to profound transformations in these and other aspects of the computing ecosystem.

Cloud is driven, first and foremost, by a transformation in demand. It is no accident that the first successful infrastructure-as-a-service business emerged from an ecommerce provider. As Amazon CTO Werner Vogels tells the story, Amazon realized, after its first dramatic expansion, that it was building out literally hundreds of similar work-unit computing systems to support the different services that contributed to Amazon's online ecommerce platform. Each such system needed to be able to scale rapidly its capacity to queue requests, store data, and acquire computers for data processing. Refactoring across the different services produced Amazon's Simple Queue Service, Simple Storage Service, and Elastic Computing Cloud, as well as other subsequent offerings, collectively known as Amazon Web Services. Those services have in turn been successful in the marketplace because many other ecommerce businesses need similar capabilities, whether to host simple ecommerce sites or to provide more sophisticated services such as video on demand.

Cloud is also enabled by a transformation in transmission. While the US and Europe still lag behind broadband leaders such as South Korea and Japan, the number of households with megabits per second or faster connections is large and growing. One consequence is increased demand for data-intensive services such as Netflix's video on demand and Animoto's video rendering—both hosted on Amazon Web Services. Another is that businesses feel increasingly able to outsource business processes such as email, customer relationship management, and accounting to software-as-a-service (SaaS) vendors.

Finally, cloud is enabled by a transformation in supply. Both IaaS vendors and companies offering consumer-facing services (e.g., search: Google, auctions: eBay, social networking: Facebook, Twitter) require enormous quantities of computing and storage. Leveraging advances in commodity computer technologies, these and other companies have learned how to meet those needs cost effectively within enormous data centers themselves [29] or, alternatively, have outsourced this aspect of their business to IaaS vendors. The commoditization of virtualization [27, 144] has facilitated this transformation, making it far easier than before to allocate computing resources on demand, with a precisely defined software stack installed.

As this brief discussion suggests, much of the innovation in cloud has occurred in areas orthogonal to the topics on which Grid computing focused—in particular, in the area of massive scale out on the supply side. The area where the greatest overlap of concerns occurs is within the enterprise, where indeed what used to be "enterprise Grids" are now named "private clouds," with the principal difference being the use of virtualization to facilitate dynamic resource provisioning.

Access to IaaS is typically provided via different interfaces from those used in Grid, including SOAP-based Web Services interfaces as well as those following the REST architectural design approach. As yet, no standard IaaS interface exists. However, Amazon's significant market share has resulted in the EC2 REST interfaces becoming almost a de facto standard. Tools such as Eucalyptus [129], Nimbus [104], and OpenNebula [150] provide access to computing resources via the EC2 interface model.

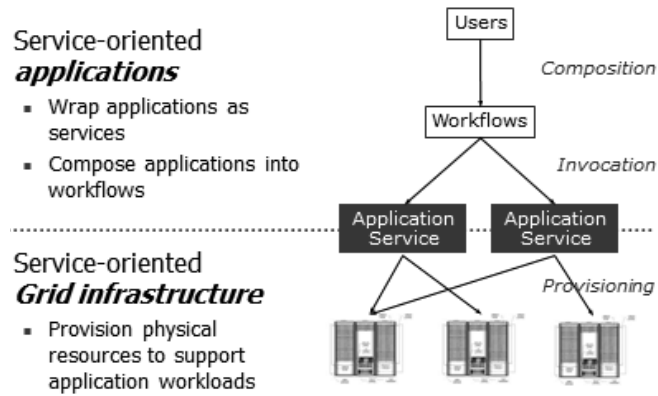

Figure 5: Use of Grid technologies for SaaS and IaaS. (From a 2004 slide by the authors.)

The emergence of cloud, and in particular IaaS, creates a significant opportunity for Grid applications and environments. During the transition of Grid middleware into a service-oriented architecture, a great deal of discussion centered on how execution management services such as GRAM could be generalized to service deployment services. Figure 5 (from 2004) shows a perspective of Grid middleware that reflects this structure. IaaS is a means of implementing a service-oriented deployment service, and as such is consistent with the Grid paradigm.

## 11. Summary and future work

Technology pundit George Gilder remarked in 2000 [90] that "when the network is as fast as the computer's internal links, the machine disintegrates across the net into a set of special purpose appliances." It is this disintegration that underpins the Grid (and, more recently, the cloud) vision of on-demand, elastic access to computer power. High-speed networks also allow for the aggregation of resources from many distributed locations, often within the contexts of virtual organizations; this aggregation has proved to be equally or even more important for many users. Whether outsourcing or aggregating, technologies are needed to overcome the barriers of resource, protocol, and policy heterogeneity that are inevitable in any distributed system. Grid technologies have been developed to answer this need.

More than 15 years of Grid research, development, deployment, and application have produced many successes. Large-scale operational Grid deployments have delivered billions of node hours and petabytes of data to research in fields as diverse as particle physics, biomedicine, climate change, astronomy, and neuroscience. Many computer science researchers became engaged in challenging real-world problems, in ways that have surely benefited the field. Investment in continued creation of Grid infrastructure continues, for example the National Science Foundation's CIF21 [2], with a focus on sustainability, virtual organizations, and broadening access and sharing of data. From the perspective of a scientist, it is hard to argue with the impact of the Grid.

While Grid-enabled applications have been limited mostly to research, the services developed to support those applications have seen extensive use. Protocols, software, and services for security, data movement and management, job submission and management, system monitoring, and other purposes have been used extensively both individually and within higher-level tools and solutions. Many of these services can now be found in one form or another in today's large-scale cloud services.

Grid computing has declined in popularity as a search term on Google since its peak around 2005. However, the needs that Grid computing was designed to address—on-demand computing, resource federation, virtual organizations—continue to grow in importance, pursued by many, albeit increasingly often under other names. In these concluding remarks, we discuss briefly a few recent developments that we find interesting.

Many years of experience with increasingly ambitious Grid deployments show that it is now feasible for research communities to establish sophisticated resource federation, on-demand computing, and collaboration infrastructures. However, obstacles do remain. One obstacle is the relatively high cost associated with instantiating and operating the services that underpin such infrastructures; a consequence of these costs is that it is mostly big science projects that make use of them. One promising solution to this problem, we believe, is to make enabling services available as hosted software as a service (SaaS) rather than as applications that must be downloaded, installed, and operated by the consumer. Globus Online [68] is an early example of the SaaS approach to VO services, addressing user profile management and data movement in its first instantiation. HUBzero [4] is another example, focused on access to scientific application software. A more comprehensive set of SaaS services could address areas such as group management, computation, research data management.

A more subtle obstacle to large-scale resource federation is that people are often unmotivated to share resources with others. The emergence of commercial IaaS providers is one solution to this obstacle: if computing power can be obtained for a modest fee, the imperative to pool computing power across institutional boundaries is reduced. Yet the need for remote access to other resources, in particular data, remains. Enabling greater sharing will require progress in policy, incentives, and perhaps also technology—for example, to track usage of data produced by others, so that data providers can be rewarded, via fame or fortune depending on context.

A third obstacle to more extensive use of Grid technologies relates to usability and scope. A Grid is still, for many, a complex service that must be invoked using special interfaces and methods and that addresses only a subset of their information technology needs. To accelerate discovery on a far larger scale, we need to address many more of the time-consuming tasks that dominate researcher time, and do so in a way that integrates naturally with the research environment. For example, research data management functions have become, with new instrumentation, increasingly time consuming. Why not move responsibility for these functions from the user to the Grid? If we can then integrate those functions with the user's native environment as naturally as DropBox integrates file sharing, we will reach many more users. With infrastructure as a service finally available on a large scale, it may well be time to move to the next stage in the Grid vision, and seek to automate other yet more challenging tasks.

## Acknowledgments

We have reported here on just a few highlights of more than 20 years of research, development, deployment, and application work that has involved many thousands of people worldwide. Inevitably we have omitted many significant results, due variously to lack of space, lack of knowledge, poor memories, or personal prejudices. We thank Sebastien Goasguen and Dan Katz for their comments, and Gail Pieper for her expert copyediting. We welcome feedback so that we can improve this article in future revisions. This work was supported in part by the U.S. Department of Energy, under Contract No. DE-AC02-06CH11357, and the National Science Foundation, under contract OCI-534113.